\documentclass[11pt]{article}
\usepackage{jheppub}
\usepackage{amsfonts,ulem,bbold}
\usepackage{amsmath,amssymb, braket}
\newcommand{\md}{\mathrm{d}}
\newcommand{\beqn}{\begin{eqnarray}}
\newcommand{\eeqn}{\end{eqnarray}}
\newcommand{\dd}{\mathrm{d}}
\newcommand{\nn}{\nonumber}

\newcommand{\be}{\begin{equation}}
\newcommand{\ee}{\end{equation}}
\newcommand{\bea}{\begin{eqnarray}}
\newcommand{\eea}{\end{eqnarray}}
\newcommand{\p}{\partial}

\newcommand{\gmn}{g_{\mu\nu}}

\newcommand{\tr}{\mathrm{Tr}}

\title{Proof of Consistency of Nonlinear Massive Gravity in 
the St\"uckelberg Formulation}
\author{S.F.~Hassan,}
\author{Angnis~Schmidt-May}
\author{and Mikael~von~Strauss}

\affiliation{Department of Physics \& 
        The Oskar Klein Centre,\\
        Stockholm University, AlbaNova University Centre, 
        SE-106 91 Stockholm, Sweden}

\emailAdd{fawad@fysik.su.se}
\emailAdd{angnis.schmidt-may@fysik.su.se}
\emailAdd{mvs@fysik.su.se}

\abstract{We address some recent concerns about the absence of the
  Boulware-Deser ghost in the St\"uckelberg formulation of nonlinear
  massive gravity. First we provide general arguments for why any
  ghost analysis in the St\"uckelberg formulation has to agree with
  existing consistency proofs that have been carried out without using
  St\"uckelberg fields. We then demonstrate the absence of the ghost
  at the completely nonlinear level in the St\"uckelberg formulation
  of the minimal massive gravity action. The constraint that removes
  the ghost field and the associated secondary constraint that
  eliminates its conjugate momentum are computed explicitly,
  confirming the consistency of the theory in the St\"uckelberg
  formulation.} 


\preprint{}
\toccontinuoustrue

\begin{document} 
\maketitle
\flushbottom

\section{Introduction and statement of the problem}

General relativity describes a massless spin-2 field $g_{\mu\nu}$. 
Constructing a mass term for $g_{\mu\nu}$ which is reparametrization
invariant requires introducing another ``metric'', say $f_{\mu\nu}$.
Then a potential can be constructed as a function of $g^{\mu\lambda}
f_{\lambda\nu}$, and one can write a massive gravity action in the
form, 
\be
S=M_p^2\int d^4x \sqrt{-g}\left[R(g) -2m^2\, V(g^{-1}f)\right]\,. 
\label{actionV}
\ee
Alternatively, this describes a massive spin-2 field in fixed
background metric $f_{\mu\nu}$.    

It has long been known that for a generic $V$ such a theory is
inconsistent, containing the Boulware-Deser ghost
\cite{Boulware:1972zf, Boulware:1973my}. Only recently potentially
consistent massive gravity theories were proposed mostly based on a
perturbative analysis in the St\"uckelberg formulation of the theory
\cite{deRham:2010ik, deRham:2010kj}. These were then generalized and
shown to be ghost free in a completely nonlinear analysis without
using the St\"uckelberg formalism \cite{Hassan:2011vm,Hassan:2011hr,
  Hassan:2011tf,Hassan:2011ea}. In spite of these proofs, concerns
about the consistency of massive gravity in the St\"uckelberg 
formulation has arisen, both at the perturbative \cite{Alberte:2010qb,
  Chamseddine:2011mu} as well as nonlinear levels \cite{Kluson:2011rt,
  Kluson:2012gz}. At the perturbative level the problem has been
addressed in \cite{deRham:2011rn,Mirbabayi:2011aa,
  deRham:2011qq}. Here these issues will be addressed and resolved at  
the completely nonlinear level.

We start with stating the problem. The simplest choice for
$f_{\mu\nu}$ in (\ref{actionV}) is the flat metric,  
\be
f_{\mu\nu}=\frac{\p \phi^A}{\p x^\mu}\eta_{AB}\frac{\p \phi^B}{\p x^\nu}\,.
\label{flatf}
\ee
The four scalar fields $\phi^A$ ensure reparametrization invariance of
$V(g^{-1}f)$ with flat $f$ and can be gauged away by a
reparametrization $\tilde x^A=\phi^A(x)$ to set $\tilde
f_{\mu\nu}=\eta_{\mu\nu}$. These are the St\"uckelberg fields that
inherit their dynamics from the form of $V$. It was pointed out in
\cite{ArkaniHamed:2002sp, Creminelli:2005qk} that the consistency of
the $\phi^A$ theory is correlated to the existence of the
Boulware-Deser ghost. This allowed for a simpler analysis of the ghost
which was otherwise an involved problem. In particular, one could work
in a ``decoupling limit'' that ignores the nonlinear dynamics of
$g_{\mu\nu}$ and retains only the $\phi^A$. By the argument of
\cite{ArkaniHamed:2002sp}, this retains also a part of the ghost
information that survives the decoupling limit. 
The potentially ghost-free potential $V(g^{-1}f)$ based on
(\ref{flatf}) was first constructed \cite{deRham:2010ik,
  deRham:2010kj} using a perturbative analysis of the St\"uckelberg
fields in the decoupling limit. It was established to be ghost
free to linear order in $h_{\mu\nu}=g_{\mu\nu} -
\eta_{\mu\nu}$. A fourth order analysis was also carried out
\cite{deRham:2010kj}.

The subsequent development in the field did not rely on the
St\"uckelberg formulation. The nonlinear demonstration of the absence
of ghost in the flat $f$ theory of \cite{deRham:2010kj} was carried
out in the unitary gauge $f=\eta$ \cite{Hassan:2011hr}. Subsequently,
these actions were extended, first, to any arbitrary non-dynamical
$f_{\mu\nu}$ \cite{Hassan:2011tf, Hassan:2011ea}, and then to a
dynamical $f_{\mu\nu}$ \cite{Hassan:2011zd, Hassan:2011ea} and shown
to be ghost free at the complete nonlinear level, without a need for
gauge fixing. This established that such theories contained the right
constraints to eliminate a propagating ghost. 

On the other hand, while the St\"uckelberg analysis, based on studying
the dynamics of the $\phi^A$ (\ref{flatf}), proved very powerful in
the decoupling limit, extending it beyond this limit led to some
speculation that the ghost may return at higher orders in $h_{\mu\nu}$
\cite{Alberte:2010qb, Chamseddine:2011mu}. This concern was addressed
in \cite{deRham:2011rn} where it was asserted that the constraints
that eliminated the ghost \cite{Hassan:2011hr, Hassan:2011tf,
  Hassan:2011zd, Hassan:2011ea} must also arise in the St\"uckelberg
formulation of the same theory. This was demonstrated perturbatively
at lowest orders in $h_{\mu\nu}$ (as well as exactly in $2$
dimensions) and argued to be extendable to higher orders. Recently, a
nonlinear analysis of constraints in the St\"uckelberg setup has been
performed \cite{Kluson:2011rt, Kluson:2012gz} and seems to indicate
that the required constraints may not exist in the St\"uckelberg
formulation. \footnote{This is different from the objection raised in
  \cite{Kluson:2011qe} and addressed in \cite{Hassan:2011ea}.} If
true, this would be in contradiction with the proofs of
\cite{Hassan:2011hr, Hassan:2011tf, Hassan:2011zd, Hassan:2011ea} as
well as the arguments in \cite{deRham:2011rn}.

Due to these lingering doubts about the consistency of massive gravity 
in the St\"uckelberg formulations and in order to conclusively 
resolve this confusion, we reconsider this set up. Our main results
are summarized below. 
\begin{itemize}
\item
With simple arguments based on general covariance, we show that
the St\"uckelberg formulation of massive gravity must agree with 
the existing consistency proofs already obtained in the standard
formulation.    
\item
We perform a Hamiltonian (ADM) analysis of the minimal model of
massive gravity in the St\"uckelberg formulation and explicitly obtain
2 constraints that remove both the ghost and its canonical
momentum. The analysis is performed at the fully nonlinear level and
for arbitrary metric $f_{\mu\nu}$. Our results extend those of
\cite{deRham:2011rn} which, for a flat $f_{\mu\nu}$, found the
corresponding constraints exactly in 2 dimensions and perturbatively
in 4 dimensions. This explicit nonlinear proof should settle the issue
of absence of ghost in the St\"uckelberg formulation of massive
gravity. 
\end{itemize}
In section 2, we discuss the consistency between the St\"uckelberg and
standard non-St\"uckelberg formulations of massive gravity.  In
section 3, we perform the consistency analysis directly in the
St\"uckelberg formalism and prove the absence of ghost at the
nonlinear level.

\section{Consistency of the St\"uckelberg and non-St\"uckelberg
  formulations}

The absence of ghost in massive gravity has been proven in the
standard, non-St\"uckelberg, formulation of the theory
\cite{Hassan:2011hr, Hassan:2011tf, Hassan:2011ea}. It is easy to
show, on general grounds, that this also implies the absence of ghost
in the St\"uckelberg formulation.

In the standard formulation, the $f_{\mu\nu}$ in (\ref{actionV}) is 
a non-dynamical tensor that ensures general covariance, while the
equations of motion are obtained by varying the action with respect to
$g_{\mu\nu}$ alone,
\be
R_{\mu\nu}-\tfrac1{2}g_{\mu\nu}R 
-2m^2V_{\mu\nu}=0\,,
\label{geom}
\ee
where, $V_{\mu\nu}=\frac{\delta V}{\delta g^{\mu\nu}}-
\tfrac1{2}g_{\mu\nu}V$. The Bianchi identity for the Einstein tensor
then implies,
\be
\nabla^\mu V_{\mu\nu}=0\,.
\label{B-const}
\ee
On the other hand, in the St\"uckelberg formulation of
(\ref{actionV}), one writes \cite{ArkaniHamed:2002sp},  
\be
f_{\mu\nu}=\frac{\p \phi^A}{\p x^\mu}\,\bar f_{AB}(\phi)\, 
\frac{\p \phi^B}{\p x^\nu}\,. 
\label{stuckelberg}
\ee
$\bar f_{AB}$ are given scalar functions of $x^\mu$, but the
St\"uckelberg fields $\phi^A$ are now treated as dynamical. Hence 
along with the $g_{\mu\nu}$ equation above, one also obtains the
$\phi^A$ equations of motion, 
\be
\frac{\delta V}{\delta\phi^A}=0\,.
\label{phieom}
\ee 
However, general covariance implies that (\ref{phieom}) is already
contained in the $g_{\mu\nu}$ equations (\ref{B-const}). To see this,
consider infinitesimal reparametrizations $\delta x^\mu=\xi^\mu$ under
which, $\delta g^{\mu\nu}=-2\nabla^{(\mu} \xi^{\nu)}$ and
$\delta\phi^A = -\xi^\mu\p_\mu\phi^A$. The invariance of the
action implies,
\begin{align}
&\delta S=-M_p^2\int\dd^4x\sqrt{g}\left[\left(
R_{\mu\nu}-\tfrac1{2}g_{\mu\nu}R 
-2m^2V_{\mu\nu}\right)\delta g^{\mu\nu}
+2m^2\frac{\delta V}{\delta\phi^A}\delta\phi^A\right]=0\,.
\end{align}
For the above variations, integrating by parts and using the Bianchi
identity then gives, 
\beqn
\nabla^\mu V_{\mu\nu} = \frac{\delta V}{\delta\phi^A}\p_\nu\phi^A\,.
\label{gct}
\eeqn
Since $\phi^A$ are non-singular coordinate transformations of $x^\mu$,
$\p_\mu\phi^A$ is an invertible matrix. Then (\ref{gct}) implies that
(\ref{phieom}) is equivalent to (\ref{B-const}), as can be explicitly
checked \cite{Hassan:2011vm}. This is a consequence of general covariance and
shows that the St\"uckelberg fields do not lead to extra equations beyond
the metric equations of motion. Adding matter does not change this
argument.

Now, the proof of absence of ghost in massive gravity in
\cite{Hassan:2011hr, Hassan:2011tf, Hassan:2011ea} only involves the
$g_{\mu\nu}$ equations of motion. In particular, it does not involve
the $f_{\mu\nu}$ equations nor gauge fixing (except for the unitary
gauge analysis in \cite{Hassan:2011hr}). Therefore, in the final
action for the five physical components of the massive spin-2 field,
obtained on imposing the constraints and eliminating the ghost, one
may express $f$ as in (\ref{stuckelberg}) and obtain the $\phi$
equations. But as discussed above, these are already contained in the
$g$ equations, showing that the St\"uckelberg formalism cannot be
inconsistent with the ghost analysis of \cite{Hassan:2011hr,
  Hassan:2011tf, Hassan:2011ea}.

After this general argument, let us briefly review the ghost
analysis for the simplest of the massive gravity actions, the minimal
model of \cite{Hassan:2011vm}, following \cite{Hassan:2011tf, Hassan:2011ea},
\be
S=M_p^2\int d^4x \sqrt{-g}\left[R(g) -2m^2\, \tr(\sqrt{g^{-1}f})\right]\,. 
\label{actionMin}
\ee  
Here, $\sqrt{g^{-1}f}$ is a square-root matrix defined such that 
$\sqrt{E}\sqrt{E}=E$. To obtain the physical degrees of freedom one
uses the ADM parametrization, 
\be
\gmn=\begin{pmatrix}
-N^2+N^i\gamma_{ij}N^j & N_i\\
N_j& \gamma_{ij}
\end{pmatrix}\,,\quad 
g^{\mu\nu}=\frac{1}{N^2}\begin{pmatrix}
-1& N^i\\
N^j& N^2\gamma^{ij}-N^iN^j
\end{pmatrix}\,.
\label{gADM}
\ee
and writes the action in the Hamiltonian formulation,
\be
S=M_p^2\int d^4x\left[\pi^{ij}\dot\gamma_{ij}+NR_0 +N^iR_i
-2m^2\, N\sqrt{\det\gamma}\,V_{min}(N, N^i,\gamma, f)
\right]\,. 
\label{actionMinADM}
\ee  
where, $V_{min}$ stands for $\tr(\sqrt{g^{-1}f})$ expressed in the ADM
parameterization and,
\be\label{GRham}
R_0=\sqrt{\gamma}\left[^3R+\left(\tfrac{1}{2}\pi^2-\pi^{ij}\pi_{ij}
\right)/\sqrt{\gamma}\right]\,,\quad
R_i=2\sqrt{\gamma}~\nabla_j\left(\pi^{ij}/\sqrt{\gamma}
\right) \,. 
\ee
In the Hamiltonian formulation, the action contains
6 potentially propagating modes $\gamma_{ij}$ and their canonically
conjugate momenta $\pi^{ij}=\delta S/\delta \dot{\gamma}_{ij}$. Of these, 5
conjugate pairs describe the massive spin-2 graviton, while the sixth
one is the Boulware-Deser ghost. The $N$ and $N_i$ have no canonical
momenta and are non-propagating. All these fields give rise to
equations of motion. The crucial point is to show that these equations
contain the two constraints that can eliminate the ghost and its
conjugate momentum. There are no equations of motion for the spectator
fields $f_{\mu\nu}$.

Of the four $N$ and $N^i$ equations of motion, three combinations
determine the $N^i$ in terms of $N$, $\gamma_{ij}$ and $\pi^{ij}$ (the
explicit solutions are obtained for three functions $n^i(N,N^i)$ in
terms of $\gamma_{ij}$ and $\pi^{ij}$). The fourth combination becomes
the Hamiltonian constraint on $\gamma_{ij}$ and  $\pi^{ij}$, 
\be
{\cal C}(\gamma_{ij},\pi^{ij})=0\,,
\ee
where $\cal C$ does not involve time-derivatives of its arguments. The 
preservation of this constraint by the time evolution of the system
requires $\mathrm{d}{\cal C}/{\mathrm{d} t}=0$. Eliminating all time
derivatives of fields using their dynamical equations leads to a
second constraint,     
\be
{\cal C}_2(\gamma_{ij},\pi^{ij})=0\,.
\ee
These two constraints eliminate the ghost field and its conjugate
momentum. Finally $N$ is eliminated by the equation following from the
preservation of ${\cal C}_2$ in time. All these equations are part of
the $g_{\mu\nu}$ equations of motion and leave behind a theory for
the 5 physical modes of a massive spin-2 field.  

The final theory also contains $f_{\mu\nu}$ which was treated as a
spectator all along. If this is now expressed in terms of the
St\"uckelberg fields $\phi^A$ (\ref{stuckelberg}), we know that
their equations of motion are already part of the $g_{\mu\nu}$
equations and there is no inconsistency. In addition, general
covariance should be used to eliminate four gauge modes. Below, we
perform a Hamiltonian analysis directly in the St\"uckelberg
formulation of the theory. 
 
\section{ADM analysis in the St\"{u}ckelberg formulation}

In this section we will perform a Hamiltonian analysis of massive
gravity in the St\"uckelberg formulation and derive the constraints
needed to eliminate the ghost. This explicitly answers the concerns
raised in \cite{Alberte:2010qb, Chamseddine:2011mu, Kluson:2011rt,
  Kluson:2012gz}, about the consistency of massive gravity in the
St\"uckelberg formulation and extends the perturbative arguments of
\cite{deRham:2011rn} to the completely nonlinear theory.

\subsection{Hamiltonian form of the action with St\"uckelberg fields} 

Here we explicitly focus on the minimal massive gravity action
(\ref{actionMin}) in the St\"uckelberg formulation with $f_{\mu\nu}$ given by 
(\ref{stuckelberg}) and reconsider the analysis of this model
in \cite{Kluson:2011rt,Kluson:2012gz}. To avoid working with the
square-root matrix, we follow \cite{Golovnev:2011aa,Buchbinder:2012wb}
to recast the action as,
\be
S=M_\mathrm{P}^2\int \dd^4 x\sqrt{g}\left[R(g) - m^2\left[{\Phi^A}_A
  +{(\Phi^{-1})^A}_B {{\bf A}^B}_A\right]\right]\,, 
\label{actionMinPhi}
\ee
where, 
\be
{{\bf A}^B}_A\equiv\partial_\mu\phi^B\,g^{\mu\nu}\,\partial_\nu\phi^C 
\,\bar{f}_{CA}\,.
\label{A}
\ee 
Note that since ${\bf A}^{AB}\equiv {{\bf A}^A}_B \bar{f}^{AB}$ is
symmetric, only the symmetric piece of $\Phi_{AB}\equiv
\bar{f}_{AC}{\Phi^C}_B$ appears in the action. We therefore treat
$\Phi_{AB}$ as symmetric in the following. Solving the ${\Phi^A}_B$
equation of motion and plugging the solution back into the action
gives back (\ref{actionMin}).


To rewrite the action (\ref{actionMinPhi}) in the Hamiltonian
formulation, we summarize the analysis of
\cite{Kluson:2011rt,Kluson:2012gz} until our conclusions diverge, and
mainly adapt their notation to facilitate comparison. In the ADM
parametrization (\ref{gADM}), the matrix ${{\bf A}^B}_A$ of (\ref{A})
can be written as, 
\be
{{\bf A}^B}_A=-\nabla_n\phi^B\nabla_n\phi^C\bar{f}_{CA}+{{V}^B}_A\,,
\ee
where,
\be
\nabla_n\phi^A\equiv\frac{1}{N}(\partial_0\phi^A-N^i\partial_i\phi^A)\,,
\quad\quad 
{V^B}_A\equiv \gamma^{ij}\partial_i\phi^B\partial_j\phi^C\bar{f}_{CA}\,.
\label{V}
\ee
In this notation, the canonical momentum conjugate to $\phi^B$ is,
\be
p_B=2m^2M_\mathrm{P}^2\sqrt{\gamma}{(\Phi^{-1})}_{AB}\nabla_n\phi^A\,.
\ee
Hence, the action in the Hamiltonian formulation becomes, \footnote{To
avoid lengthy expressions, we mostly work with the action principle
in the Hamiltonian formulation. Only in section 3 we work with Poisson
brackets in a minimal way.}
\be
S=M_p^2\int d^4x\left[\pi^{ij}\dot\gamma_{ij}+p_A\dot\phi^A-{\cal H}
\right]\,, 
\label{actionMinADM2}
\ee  
with the Hamiltonian density given by,
\be
\mathcal{H}=-NR_0-N^iR_i+ N\mathcal{H}_T^\mathrm{sc}+
N^i\mathcal{H}^\mathrm{sc}_i\,.
\label{H}
\ee
The first two terms are familiar from general relativity, whereas the
rest read as,   
\begin{align}
\mathcal{H}_T^\mathrm{sc}&=\sqrt{\gamma}M_\mathrm{P}^2m^2
\big({\Phi^A}_A+{(\Phi^{-1})^A}_B{V^B}_A\big)+
\frac{1}{4m^2M_\mathrm{P}^2\sqrt{\gamma}}\,\Phi^{AB}p_A\,p_B\,,\nn\\
\mathcal{H}^\mathrm{sc}_i&= p_A\,\partial_i\phi^A\,.
\end{align}

Before we set out to determine the physical content of the theory, 
let us begin by listing its field content in the Hamiltonian
description:  
\begin{table}[htdp]
\begin{center}
\begin{tabular}{|c|c|c|}
\hline
fields & components & canonical momenta\\
\hline
$N$& 1&-\\
$N_i$& 3&-\\
$\gamma_{ij}$& 6&$\pi^{ij}$\\
$\phi^A$ &4 & $p_A$ \\
${\Phi^A}_B$ & 10&-\\
\hline
\end{tabular}
\end{center}
\end{table}%
\\
Thus there are 10 potentially propagating modes given by $\gamma_{ij}$
and the St\"uckelberg fields $\phi^A$, while 14 further fields are
non-propagating. There are also 4 gauge invariances.

A naive counting of the field components and non-dynamical equations
of motion (ones without time derivatives) derived from
(\ref{actionMinADM2}) may give the impression that the theory contains
the Boulware-Deser ghost: At first sight, the 10 equations of motion
for the ${\Phi^A}_B$ depend on ${\Phi^A}_B$ and will therefore
determine this matrix rather than serve as constraints on other
fields. Since the action is linear in $N$ and $N^i$, their
equations of motion turn into 4 non-dynamical equations for the
remaining variables. But unlike general relativity, now these also
contain the $p_A$ and $\phi^A$ and can be solved, for example, for the
$p_A$ (after gauge fixing the $\phi^A$) rather than impose a
constraint on the Boulware-Deser ghost contained in
$\gamma_{ij}$. Hence, it seems that the theory may not have the
required constraints to reduce the number of propagating modes of
$\gamma_{ij}$ below 6, as concluded in \cite{Kluson:2011rt,
  Kluson:2012gz}. 

However, as we will demonstrate in the following, the 10
equations of motion for ${\Phi^A}_B$ depend only on 9 independent
combinations of the matrix elements~${\Phi^A}_B$. One combination
remains undetermined and the corresponding equation of motion will
give an additional constraint on the remaining variables instead,
which will remove the ghost. Below we obtain this and the associated
secondary constraint. 

\subsection{The first constraint}

Varying the action (\ref{actionMinADM2}) with respect to ${\Phi^{AB}}$
gives the equations of motion \cite{Kluson:2012gz}, 
\be\label{eomPhi}
\Psi_{AB}\equiv\sqrt{\gamma}M_\mathrm{P}^2m^2\Big(\bar{f}_{AB}-
(\Phi^{-1})_{AC}V^{CD}(\Phi^{-1})_{DB}\Big)+
\frac{1}{4m^2M_\mathrm{P}^2\sqrt{\gamma}}\,p_A\,p_B=0\,.
\ee
To arrive at these equations we have divided by $N$ which is non-zero
since $\gmn$ is invertible. Since $\Psi_{AB}$ is symmetric, the naive
expectation would be that (\ref{eomPhi}) provides 10 conditions that
determine all 10 components of the symmetric $(\Phi^{-1})^{AB}$. 

However, the crucial observation 
is that the $4\times 4$ matrix 
$V^{AB}\equiv\gamma^{ij}\partial_i \phi^A\partial_j\phi^B$ always has
rank 3 since it is composed of $3\times 4$ and $3 \times 3$ matrices 
each of maximum rank 3. This also implies that
$W\equiv\Phi^{-1}V\Phi^{-1}$ has rank 3 and hence cannot depend on
more than 9 independent combination of the 10 $\Phi_{AB}$. To see
this, note that, being a symmetric rank-3 matrix, $W$ can be
diagonalized by an appropriate orthogonal transformation $O$ to,
\be
W_\mathrm{D}=O^\mathrm{T}WO=\text{diag}\{w_1,w_2,w_3,0\}\,,
\ee 
where $W_\mathrm{D}$ always has one zero eigenvalue. Since the
orthogonal matrix $O$ has six independent parameters, on inverting the
transformation we see that $W$ depends on at most 9 independent
parameters. This shows that the 10 equations (\ref{eomPhi}) depend
only on the 9 independent combinations of $\Phi^{AB}$ that appear in
$\Phi^{-1}V\Phi^{-1}$ and can only determine these combinations. One
combination of the equations therefore cannot fix $\Phi^{AB}$, rather 
it constrains $\gamma_{ij}$ and $p_A$.  

The constraint hidden in (\ref{eomPhi}) is extracted by multiplying on
both sides with $\Phi$ to get,  
\be\label{multeom}
\Phi^{CA}\left(\bar{f}_{AB}+\frac{1}{4(m^2M_\mathrm{P}^2\sqrt{\gamma})^2}
\,p_A\,p_B\right)\Phi^{BD}=V^{CD}\,.
\ee
As noted above, $V^{CD}$ is a rank-3 matrix and so the left-hand side
also has to have rank 3. By definition, $\Phi$ is invertible and thus
has rank 4. Therefore the matrix   
\be
{\bf C}_{AB}\equiv\bar{f}_{AB}+\frac{1}{4(m^2M_\mathrm{P}^2 
\sqrt{\gamma})^2}\,p_A\,p_B
\ee
is constrained to have rank 3 by equation (\ref{multeom}). This
implies the constraint ${\mathcal C}\equiv\det {\bf C}=0$.
Since $p_A\,p_B$ is a rank-1 matrix, $\det {\bf C}$ is particularly  
simple and the constraint becomes,
\be\label{constrainteq}
{\mathcal C}\equiv
{\bar{f}}\left(\bar{f}_{AB}\frac{p^Ap^B}{\alpha^2 \gamma} +1
\right)= 0\,, 
\ee
where $\bar{f}\equiv \det( \bar{f}_{AB})$ and $\alpha=2m^2
M_\mathrm{P}^2$. This is a single constraint on $\gamma_{ij}$ and
$p_A$ and can be used to determine $\gamma=\det(\gamma_{ij})$ in terms
of the $p_A$. Together with the equations obtained from varying the
action with respect to $N$ and $N^i$ (which can be solved for the
$p_A$ on gauge fixing the $\phi^A$ fields), this gives a single
constraint on the $\gamma_{ij}$ and the $\pi^{ij}$ that can remove the
ghost field. The associated secondary constraint will be obtained in
the next subsection. 

To compare with existing results, in \cite{deRham:2011rn} the analogue
of (\ref{constrainteq}) was obtained in 2 dimensions and also
perturbatively in 4 dimensions to linear order in
$h_{\mu\nu}=g_{\mu\nu}-\eta_{\mu\nu}$ (in this case, for a different
choice of the potential).  Here we have derived the constraint in the
minimal model of massive gravity in St\"uckelberg formulation at the
fully nonlinear level. The proof can be generalized to any dimension
and, unlike previous studies, is valid for general ${f}_{AB}$.

We now obtain a result that will be used in the next subsection. Let
us write (\ref{eomPhi}) in the form $\Phi^{CA}\Psi_{AB} p^B=0$. On
imposing ${\mathcal C}=0$ and using $V$ given in (\ref{V}) this
becomes,  
\be 
\label{zeroeigv}
V^{CD}{(\Phi^{-1})_{D}}^Ap_A=\partial_i\phi^C\gamma^{ij}
\partial_j\phi^D{(\Phi^{-1})_{D}}^Ap_A
\approx 0\,.
\ee
In words, ${(\Phi^{-1})_A}^Cp_C$ is the eigenvector with zero
eigenvalue of the rank-3 matrix ${V_B}^A$. These equations imply a
result that will be used later, 
\be 
\gamma^{kj}\partial_j\phi^D{(\Phi^{-1})_{D}}^Ap_A\approx0\,.
\label{pcontraction}
\ee
The last step can be justified explicitly. The map 
$\partial_i\phi^C$ can produce any 3-vector $B_i$ as a map from  
some 4-vector $b_C$. In particular, one can always find three
4-vectors $b^{a}_C$ $(a=1,2,3)$ that map to 3 linearly independent
3-vectors $B_i^a$, 
\beqn\label{defB}
{B_i}^a\equiv\partial_i\phi^C\,b^{a}_C\,, \qquad a=1,2,3\,.
\eeqn
Then, ${B_i}^a$ regarded as $3\times 3$ matrix is invertible (easiest  
to verify in a unitary gauge $\tilde\phi^C=x^C$). Now multiplying 
(\ref{zeroeigv}) with ${(B^{-1})^k}_a({b^a})_C$ gives
(\ref{pcontraction}). 

\subsection{The secondary constraint}

In order to eliminate both the ghost and its conjugate momentum,
a second constraint is needed. This should arise as a secondary
constraint that preserves (\ref{constrainteq}) under time evolution,  
i.e., ${\mathcal C}_2(\gamma,\pi)\propto d{\mathcal C}/dt
\approx 0$. It is most efficient to compute $d{\mathcal C}/dt$ using
Poisson brackets, 
\footnote{The Poisson bracket is defined as  
$
\{f(x), g(y)\}\equiv\int \md^3 z\left(\frac{\delta f(x)}{\delta
q^\alpha(z)}\frac{\delta g(y)}{\delta p_\alpha(z)}-\frac{\delta
f(x)}{ \delta p_\alpha(z)}\frac{\delta g(y)}{q^\alpha(z)}\right)  
$, where $q^\alpha$ and $p_\alpha$ stand for the variables $\phi^A$, 
$\gamma_{ij}$ and their canonical momenta $p_A$, $\pi^{ij}$,
respectively.} 
\be
\frac{d{\mathcal C}}{dt} =\{\mathcal{C}\,,\,H_\mathrm{tot}\} \,.
\ee
Below, we compute this and verify that it indeed leads to a
constraint.  

An economical way of going over from the action (\ref{actionMinADM2})
to the Poisson bracket formulation, is to regard only 
$(\gamma_{ij},\pi^{ij})$ and $(\phi^A,p_A)$ as canonical pairs and
obtain all non-dynamical equations in terms of Lagrange multipliers.
This requires introducing the Hamiltonian, 
\be
\label{htot2}
H_\mathrm{tot}=\int d^3 x\left[N(\mathcal{H}_T^\mathrm{sc}-R_0 +
  \Gamma^{AB}\Psi_{BA})+N^i(\mathcal{H}^\mathrm{sc}_i-R_i)\right]\,.   
\ee 
Note that as compared to (\ref{H}) this contains the extra Lagrange
multipliers $\Gamma^{AB}$ to obtain the $\Phi^{AB}$ equations of
motion $N\Psi_{BA}$ (\ref{eomPhi}). Of course, the theory can also be
  extended by introducing extra momenta and Lagrange multipliers as in
  \cite{Kluson:2012gz}. But that will not affect the computation here. The
  $\Gamma^{AB}$ are determined by the new $\Phi_{AB}$ equations of
  motion. 

To compute the Poisson brackets we need the following identities,
\begin{align}\label{pbid}
&\{p_A(y),\partial_{x^i}\phi^B(x)\}=-\delta^B_A\partial_{y^i}
\delta(x-y) \,, 
\qquad \{p_A(y), p_B(x)\}=~0\,,\nonumber\\ 
&\{\bar{f}_{BC}(y),p_A(x)\}=\frac{\delta\bar{f}_{BC}(y)}{\delta\phi^A(x)}
\,,\hspace{1.9cm}\quad\{\gamma^{-1}(y), \gamma_{ij}(x)\}=~0\,,
\end{align}
together with (where $\nabla$ is the $\gamma_{ij}$ compatible
covariant derivative),   
\begin{align}\label{bimid}
\frac{\delta R_i(x)}{\delta\pi^{jk}(z)}&=-(\gamma_{ik}(x)\nabla_{z^j} 
+\gamma_{ij}(x)\nabla_{z^k})\delta(x-z)\,,\nonumber\\ 
\frac{\delta R_0(x)}{\delta \pi^{jk}(z)}&=\frac{1}{\sqrt{\gamma(x)}}
\Big(\gamma_{jk}(x)\pi(x)-2\pi_{jk}(x)\Big)\delta(x-z)\,.
\end{align}
For $\{\mathcal{C}, H_\mathrm{tot}\}$ to give a constraint on $\gamma$
and $\pi$, it should not depend on $N, N^i$ and the single component
of $\Phi^{AB}$ not determined by (\ref{eomPhi}). Using the above
identities we find,
\be
\{\mathcal{C}(y), \mathcal{H}^\mathrm{sc}_i(x)\}\approx -2\bar{f}(y)
\nabla_{y^i}
\delta(x-y)
\approx\{\mathcal{C}(y), R_i(x)\}\,.
\ee
Hence the term proportional to the $N^i$ in $H_\mathrm{tot}$ does not
contribute to ${\mathcal C}_2$ on the constraint surface. The
remaining terms are, 
\be\label{ncontrib}
\{\mathcal{C}(y), H_\mathrm{tot}\}=\int\md^3x~N(x)
\{\,\mathcal{C}(y)\,,\, \mathcal{H}_T^\mathrm{sc}(x)+
{\Gamma^A}_B(x){\Psi^B}_{A}(x)- R_0(x)\,\}\,.
\ee
Each bracket can be evaluated on the constraint surface using
(\ref{pbid}) along with, 
\begin{align} 
\label{pbrelations}
\{p_A(y),{V^C}_D(x)\}&=-\gamma^{ij}(x)\partial_{x^i}\phi^B(x)\Big[
\big(\delta^C_A\bar{f}_{BD}+\delta^C_B\bar{f}_{AD}\big)
\partial_{y^j}+\partial_{x^j}\phi^C
\frac{\delta\bar{f}_{BD}}{\delta\phi^A}\Big]_x\delta(x-y)
\,,\nonumber\\
\{\mathcal{C}(y),R_0(x)\} &\approx \bar{f}\pi\gamma^{-1/2}
\delta(x-y) \,.
\end{align}
The coefficient of the term proportional to $\p_j N$ vanishes by
virtue of (\ref{pcontraction}). It is then straightforward to see that
the result is proportional to $N(y)$ which therefore appears as an
overall factor in $\{\mathcal{C},H_\mathrm{tot}\}$ and can be divided
out. Then the secondary constraint reads,   
\be\label{secconst}
{\mathcal C}_2=N^{-1}\{\mathcal{C}(y), H_\mathrm{tot}\}\approx
- \gamma^{-1/2}\bar{f} \pi
-4m^2 M_{\mathrm P}^2 \gamma^{-{1/2}}\bar{f}{(\Phi^{-1})}_{BD}
\partial_j\phi^D\gamma^{jk}\nabla^{(f)}_kp^B \,.
\ee
Here,$\nabla^{(f)}_kp^B=\partial_kp^B+\Gamma^B_{CD}\partial_k
\phi^Cp^D$, with $\Gamma^B_{CD}$ being the Levi-Civita connection of
the metric $\bar{f}_{AB}$. To get (\ref{secconst}), we have eliminated 
$\Gamma_{AB}$ using the $\Phi^{AB}$ equation,
\be\label{eqforgamma}
\Gamma_{AB}(\Phi^{-1})^{BC}V_{CD}\approx0\,,
\ee
and in deriving the second term we have made use of
(\ref{pcontraction}). Note that (\ref{secconst}) still contains the
$\Phi^{AB}$, but it is easy to see that it depends only on the 9
combinations of these that are determined in terms of other variables
by $\Psi_{AB}=0$: Multiplying (\ref{eomPhi}) by
${(B^{-1})^k}_a({b^a})_C\Phi^{CA}$, where ${B_k}^a$ is defined in
(\ref{defB}), we obtain,  
\be
{(B^{-1})^k}_a({b^a})_C\Phi^{CA}\left(\bar{f}_{AB}+\frac{p_Ap_B}
{\alpha^2\gamma}\right)=(\Phi^{-1})_{BD}\partial_j\phi^D\gamma^{jk}\,. 
\ee
Hence the equation $\Psi_{AB}=0$ that determines only 9 components of
$\Phi^{AB}$, contains this matrix in the same combination in which it
appears in (\ref{secconst}). Thus on gauge fixing the $\phi^A$ by using
coordinate transformations, solving for the $p^A$ using the $N$ and
$N^i$ equations and eliminating the 9 components of $\Phi$ using 
$\Psi_{AB}=0$, we obtain a constraint ${\mathcal C}_2(\gamma,\pi)
\approx 0$. This has the desired form to eliminate the momentum
canonically conjugate to the ghost field. Finally note that
(\ref{eqforgamma}) along with (\ref{zeroeigv}) implies that
$\Gamma_{AB}=\lambda p_A p_B$.


\end{document}